%
%
\input psfig.tex
\magnification=1200
\hoffset=-.25in
\voffset=.2in
%
 
\vsize=8.0in 
\hsize=6in 
\tolerance 10000 
 
\ \ 
\bigskip
\noindent {\bf } \hfill{BERC-PHY-99/102}

\vskip 1.2in
\centerline{\bf Limits to Sympathetic Evaporative Cooling of a 
Two-Component Fermi Gas}
\noindent 
\medskip        
\vskip .5in
\centerline{\it by}
\vskip .5in
\centerline{\rm M. Crescimanno, C. G. Kaoy and R. Peterson}
\bigskip
\centerline{\it Physics Department}
\centerline{\it Berea College}
\centerline{\it Berea, Ky. ~40404}
\vskip .7in
\centerline{\it August 1999}
\vskip 1.2in
{\bf ABSTRACT:} We find a limit cycle in a quasi-equilibrium model 
of evaporative cooling of a two-component trapped fermion gas. The 
existence of such a limit cycle represents an obstruction to 
reaching the quantum ground state evaporatively. We show that 
evaporatively $\beta\mu \sim {\cal O}(1)$. 
We speculate that one may be able to cool
an atomic fermi gas further 
by photoassociating dimers near the bottom of the fermi sea.

\vfill
\eject
\ \ 

\ \
\vskip .2in
\ \ 
\bigskip
\noindent {\bf I. Introduction:} The spectacular successes 
of laser cooling techniques
in creating Bose-Einstein condensation (BEC) in trapped 
dilute alkali vapors$^{[1,2,3]}$ has stimulated efforts to 
form dilute nearly degenerate atomic fermion gases.
Such systems undoubtedly have unique phenomenology, and 
since one can control the composition, densities and 
even scattering lengths in principle, they 
furnish a window to familiar phenomena (such as superconductivity, etc.)
in unusual parameter regimes.$^{[4,5,6,7,8]}$
A critical step in achieving BEC in dilute alkali vapors 
is evaporative cooling. In this note we describe limitations 
to the use of evaporative cooling for a harmonically trapped  
two-component fermi system.

Overall antisymmetry of the final state wavefunction 
forces the
s-wave scattering amplitude for two-body collisions
in a single component polarized fermi gas to vanish identically.
However, in a system composed of two or more fermi species
there can still be appreciable s-wave scattering amplitudes
at low energies.
Recent experimental observation of quantum statistical effects
reducing the scattering frequency at low temperature has been reported
in reference [9]. Sympathetic evaporative cooling in two-component 
Bose systems has been experimentally verified$^{10}$. 
Aspects of the dynamics of sympathetic cooling in a two component 
fermi system have been discussed theoretically$^{11,12}$, and 
recently achieved experimentally$^{13}$.

We will show in a robust model 
that sympathetic cooling of trapped fermions is intrinsically 
limited to 
${{\mu}\over{T}}\sim {\cal O}(1) $ where $T$ is the temperature and
$\mu$ is the chemical potential. 
Largely independent of trap and atomic parameters, this limit indicates
that sympathetic evaporative cooling
alone cannot achieve
occupation probabilities in the trap single particle ground state
characteristic of 
typical degenerate fermi systems ({\it e.g.} atomic nuclei 
and typical metals). 

In summary we model evaporation 
as a succession of quasi-equilibrium states.
We find that evaporation moves the
chemical potential towards saturation at a  
fixed (non-zero) fraction of the evaporation energy. 
Lowering the evaporation energy 
in an attempt to cool further simply causes the 
fermi surface and temperature to both recede, 
thus not substantially increasing the 
occupation probability of the lowest single particle state. 

We demonstrate this limiting behavior in a model with two regimes, one for 
which the evaporative energy scales of both fermion types 
are identical and the opposite extreme where one
selectively only allows evaporation of one species from
the trap. Rather than end on a pessimistic note, we 
conclude by speculating on a possible method for 
surmounting the difficulty of cooling an atomic fermi gas.
\bigskip
\noindent {\bf II. The Model:} We aim to explain 
general features of the evaporative 
cooling of a two-component fermion gas without recourse to the 
details of dynamics and transport. In particular, trap lifetime and 
other timescales will play almost no role in our considerations. 

As described above, we focus on the case of a two-fermion system
cooling sympathetically by evaporation through 
interspecial two-body collisions only. 
As described in the introduction, this is a reasonable
assumption as long as the density is low and the 
temperature is low enough that the amplitude for 
intraspecial scattering is small.

In the limit of interest for the first regime, we model 
the evaporation process
as a succession of quasi-equilibrium states in which the distribution 
functions of both species are cut off at $E_{evap}$. 
Aspects of the thermodynamics of dilute harmonically trapped  
fermions without an energy cut-off are described in Ref.[5]. 
Furthermore, 
we ignore the effect of any other environmental fields
(for example, trap fields), 
and assume them 
to be constant over the lifetime of the system. The average effect of the 
interactions between the different species is absorbed into a mean 
field term which we sweep into $\mu$
(see, for example, Ref.[18]). We further assume that particle 
number is not communicated between the species, 
and so their individual 
$\mu$ can differ.
Interspecial
scattering processes do communicate
energy between species, and so we assume 
that both species are always a at common
temperature

Consider evaporating both 
species at the same energy cut-off $E_{evap}$. We complete the 
analysis for this case and then turn to the opposite extreme
where only one species evaporates. 
If both fermion species $a$ and $b$ are (nearly) the same mass
then crossing symmetry equates the total rate
for scattering into final state $|a,b>$ with that of 
$|b,a>$. This means that for the same evaporative cutoff, the 
rate at which particles of one species evaporates equals that 
of the other. 

Finally, for simplicity, we model the evaporative process
as one which always reduces the particle number by one and
removes energy $E_{evap}$. 
In the model we develop  we will ignore the contribution to the cooling 
that results from the interspecial mean field$^{18}$. In the 
cases of interest in current experiments this interspecial mean field 
energy is expected to be very small compared with the other 
relevant energy scales (for example the fermi energy). 
Although this model of evaporation is a gross simplification, 
it becomes a progressively better approximation 
as the temperature drops, and we are confident it
captures the main features
of the evaporative process.

We approximate 
each component's scaled 
number and energy by cut-off equilibrium distribution 
functions
$$ N = \int_0^1 {{x^d~{\rm d} x}\over{e^{\beta(x-\mu)}+1}}
\qquad \qquad 
E =  \int_0^1 {{ x^{d+1}~{\rm d} x}\over{e^{\beta(x-\mu)}+1}}
\eqno(1)$$
where $\beta$ and $\mu$ are respectively the inverse temperature and 
the chemical potential both made 
dimensionless by factors of $E_{evap}$. 
Further, $d$ depends on the actual spectrum of the trap, and 
for a three dimensional isotropic harmonic trap is 2. 
We will keep the discussion rather 
general with respect to $d$, but use $d=2$ in 
all the graphs and particular conclusions below. While the actual 
value of $d$ does not substantially impact the nature of 
the conclusions that we draw, we do find that reducing $d$ 
(by, for example, significantly changing the aspect ratio of
the trap) makes sympathetic evaporative cooling generally less effective.  
Finally in Eq.(1) $E$ is dimensionless ({\it i.e.} in units of $E_{evap}$)
and we have suppressed some overall factors that depend on $E_{evap}$ and
trap frequency. 

We model evaporative cooling by simply
following Eq.(1) through flow along 
$$ \biggl(\matrix{ {\rm d}N \cr  
 {\rm d}E \cr 
}\biggr) 
= \biggl(\matrix{ {-1 } \cr 
{ -1 } \cr 
}\biggr) {\rm d}N
\eqno(2)$$
By using symmetries of the scattering for fermion species of 
nearly equal mass
for the case where both fermion species are
being evaporated at the same energy $E_{evap}$, we find that 
the net effect on the individual distribution functions is encapsulated
in Eq.(1) and Eq.(2) for each species separately (and so in what 
follows for this regime we suppress indices). 

The resulting differential equations for $\beta$ and 
$\mu$ along this evaporative trajectory read,
$$ {{\partial \beta}\over{\partial N}} = 
{{1}\over{det(M)}} \biggl( -{{\partial E}\over{\partial \mu}}|_\beta
+{{\partial N}\over{\partial \mu}}|_{\beta}\biggr)
\eqno(3)$$
$$ {{\partial \mu}\over{\partial N}} = {{1}\over{det(M)}}
\biggl( {{\partial E}\over{\partial \beta}}|_\mu - 
{{\partial N}\over{\partial \beta}}|_\mu\biggr)
\eqno(4)$$
where the determinant $det(M)$ is given via
$$ det(M) = {{\partial N}\over{\partial \beta}}|_\mu 
{{\partial E}\over{\partial \mu}}|_\beta - 
{{\partial N}\over{\partial \mu}}|_\beta
{{\partial E}\over{\partial \beta}}|_\mu
\eqno(5)$$
which, from the quasi-equilibrium distribution functions
Eq.(1) we find $det(M) >0$ for all $\beta$, $d$ and $\mu$. 
This positivity may be understood on general grounds via the 
connection between $det(M)$ and the specific heat, $c_V$ at
constant $N$,
$$ c_V = {{\beta^2 det(M)}\over{ {{\partial N}\over{\partial \mu}}|_\beta}}
\eqno(6)$$
and by the fact that ${{\partial N}\over{\partial \mu}}|_\beta > 0$ 
as a consequence of, for example, Eq.(1). 
The $det(M)$ vanishes 
in the low temperature limit as $\sim {{\pi^2\mu^4}\over{3\beta^3}}$. See 
Figure 1 for an example of this behavior (for $d=2$ and $\mu=2/3$). 
\medskip
\centerline{ \psfig {figure=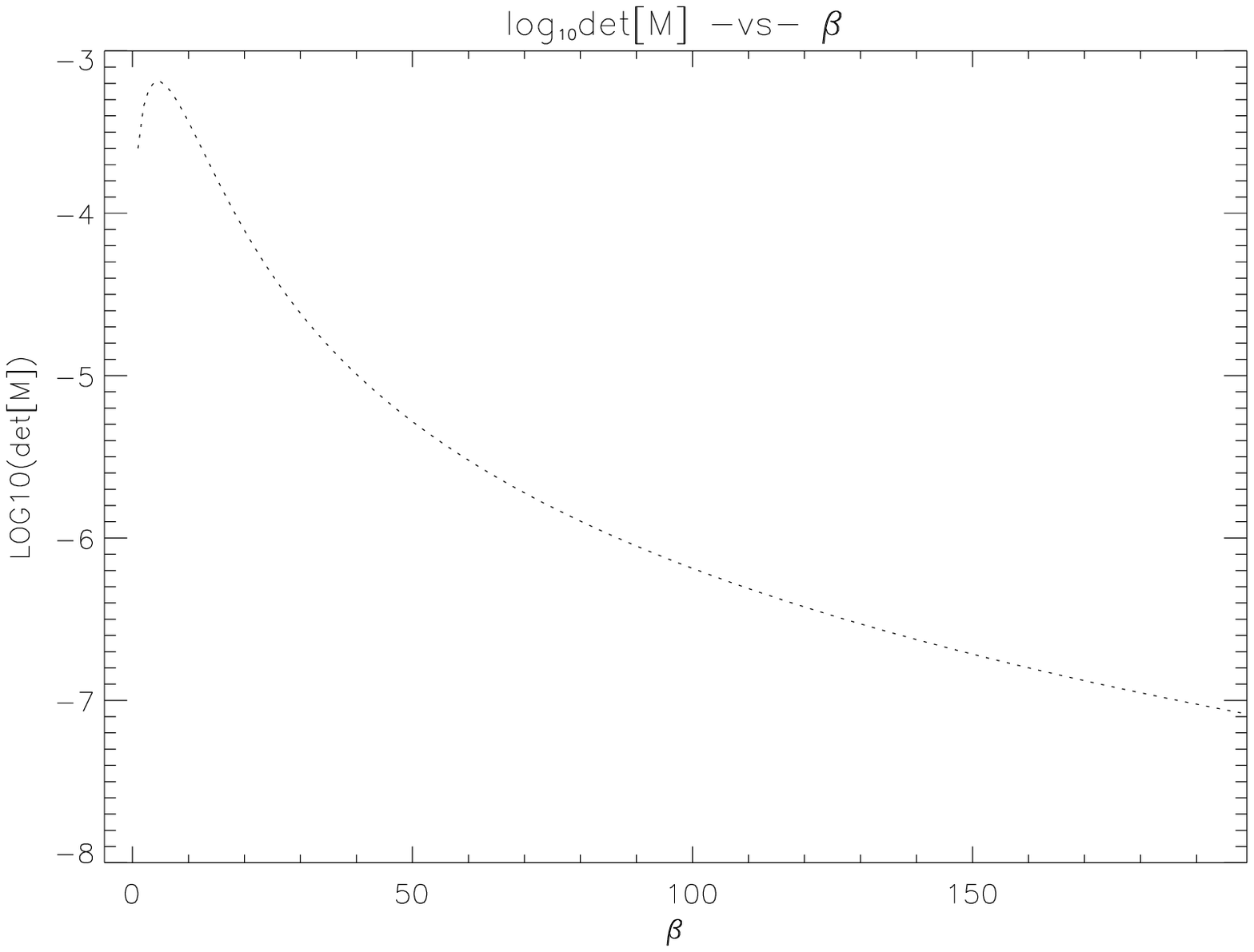,height=2.0in,angle=0}}
\smallskip
\centerline{\it Figure 1: $det(M)$ for $\mu=2/3$, as a function of $\beta$.}
\medskip
Starting far from degeneracy, the net effect of the 
evaporative process is to alter $\mu$ while increasing $\beta$. 
Note that using the equilibrium distributions Eq.(1)
implies by Eq.(3) that $\beta$ increases monotonically along the flow. 
Upon reflection we see that, $\mu$'s evolution 
does not share this property. Instead, 
we find that the system Eqs.(1), (3) and (4) has a  limit cycle at 
${{\partial E}\over{\partial \beta}}|_\mu-
{{\partial N}\over{\partial \beta}}|_\mu =0$ in the ($\beta,\mu$) plane. 
This limit cycle is at an intermediate 
value of $\mu = \mu^*(\beta)$ for all temperatures. 

Figure 2 below is a graph of $\mu^*$ as a function of temperature
for $d=2$
\medskip
\centerline{ \psfig {figure=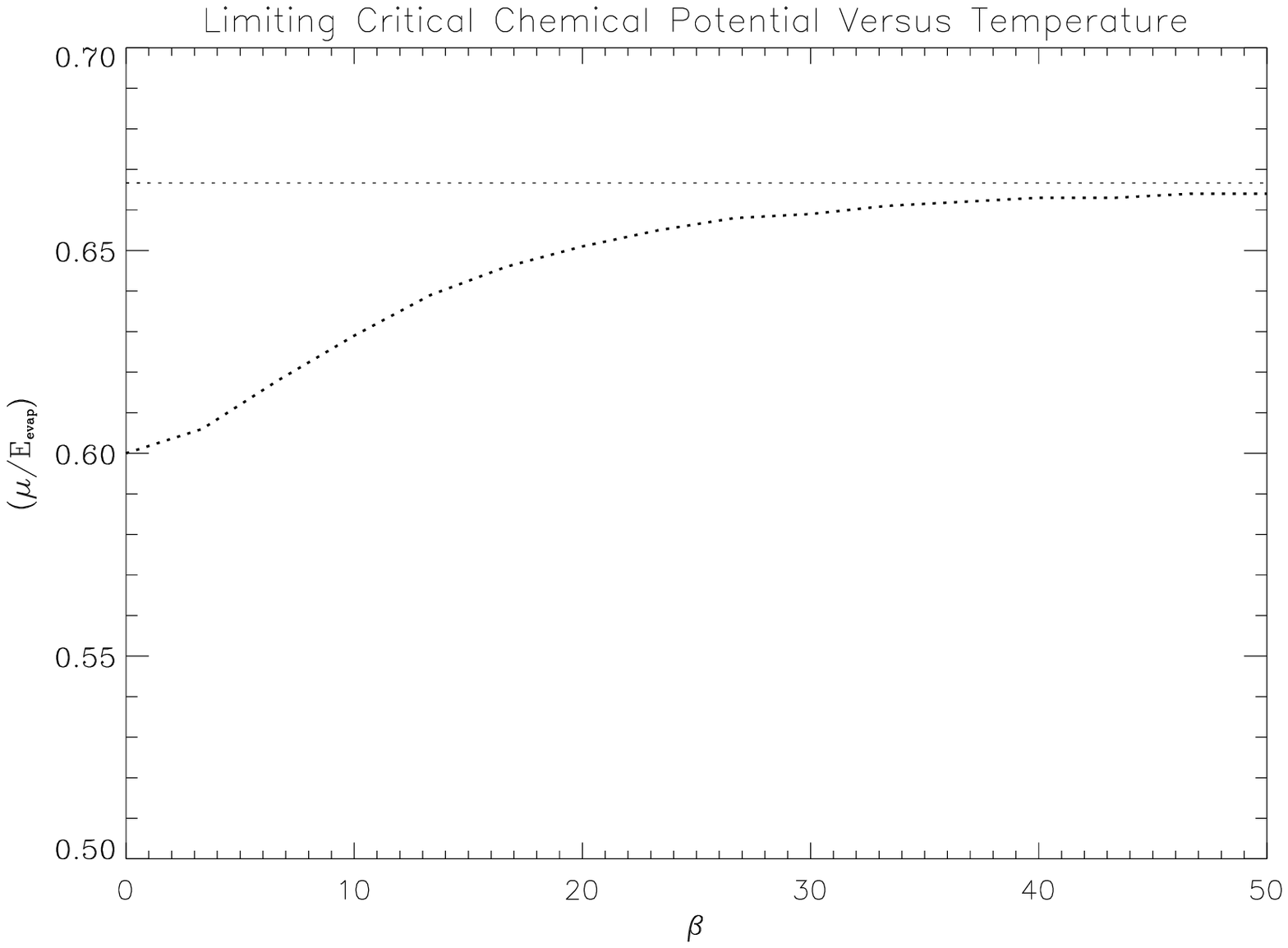,height=2.0in,angle=0}}
\centerline{\it Figure 2: The Evaporative Limit Cycle with 
the Low-Temperature Asymptote}
\medskip
For $d=2$, the low temperature limit of $\mu^*$ is $2/3$.
We reiterate,  that for values of $(\beta,\mu)$ below the curve, 
the evaporative process increases $\mu$ whereas for $(\beta,\mu)$ values 
above the curve they {\it reduce} it. 
Thus, as one tries to lower the 
temperature lower by reducing $E_{evap}$ (thus increasing $\mu$),
evaporation causes the fermi surface 
$\mu^* E_{evap}$ to also drop away. 
Of course, as one drops $E_{evap}$ the scaled inverse temperature
$\beta$ 
also drops trivially.

A figure of merit measuring how 
close one is to the quantum many body ground state is the 
occupation probability in the trap single particle ground state. 
This is a function of
the ratio of the chemical potential to the temperature, 
in our notation $\beta \mu$. This combination is independent
of $E_{evap}$ in our simple model. We now 
show that this product is limited by the total 
scaled atom number $N$ (which in the normalization of Eq.(1) is 
limited to be at most ${{1}\over{d+1}}$). 

We consider two cooling methodologies which we refer to alternatively as
``passive'' and ``active''. They refer respectively to holding the 
$E_{evap}$ fixed or suddenly reducing it. Unfortunately, in actual
experiments underway$^{13}$, $E_{evap}$ is varied continuously, so these
two cooling methodologies are probably not directly related to current 
experiments. This choice of cooling methodologies  
is advantageous analytically since it does allow 
evaluation of the 
effects of evaporation without recourse
to any dynamical timescales. 

In the passive method, the 
system is held at a fixed $E_{evap}$ and 
allowed to cool by evaporation indefinitely.
Dynamically, since the emerging fermi surface at $\mu^*$ is always
a fraction of $E_{evap}$, the cooling rate is 
limited by the overall rate of escape, which (at low temperature)
is Boltzmann suppressed
by a factor of $e^{-\beta(1-\mu*)}$, and by the trap lifetime. 
However, we reiterate the ``kinetic'' model we employ analytically 
encapsulates
limits to  sympathetic evaporative cooling 
of the two component fermion system without
including  ``dynamical'' effects (such as 
collisional or trap timescales). 

In the active method on the other hand
we abruptly
lower $E_{evap}$ to a value at or below our initial 
$\mu^* E_{evap}$. Call this new evaporation energy $E'_{evap}$.
The distribution function is invariant under such a change in $E_{evap}$. 
The $\mu$ and $\beta$ values will trivially jump by factors of the 
evaporative energy scale ratio. Of course, the overall scaled
phase space constants that we suppressed in Eq.(1) do go as 
positive powers of 
$E_{evap}$. Dropping $E_{evap}$ to $E'_{evap} < E_{evap}$ has 
the immediate effect of dropping $N$, for example. This corresponds
precisely to the statement that the all the particles with $E>E'_{evap}$ 
leave immediately. 

Since $\mu\beta$ measures our progress towards the ground state and
is invariant under a 
sudden drop in $E_{evap}$, we see that only subsequent evaporation 
(and rethermalization) of 
the remaining fermions can result in increases in the product $\mu\beta$. 
We now show that evaporation after the drop in $E_{evap}$ does not lead
the system  
substantially closer to the quantum ground state. 

To recapitulate, the relevant question in both 
cooling methodologies considered here
is whether
the subsequent evaporative cooling 
leads to a large $\beta \mu$ product. 

The evaporation equations above can be integrated 
numerically for $\beta(\mu)$ by eliminating $N$ from Eq.(3) and Eq.(4). 
In doing so, one finds that for values of 
$\mu \ne \mu^*$, the change in the dimensionless temperature 
ratio $\beta$ is generally relatively small, on the order of $\beta$ itself. 
To get to a nearly degenerate fermi system starting far away from 
the ground state, we need a cooling regime in 
which much larger temperature drops are achievable. In numerical
simulations, one finds that the only large temperature changes
happen evaporatively when the system is at a $\mu$ very near $\mu^*$, 
basically within 10 percent of that value. 
Figure 3 is the integral $\beta(\mu)$ of Eq.(3) and Eq.(4)
for two initial conditions, $(\mu_i,\beta_i) = (0,2.5)$
and $(2.8,4.5)$.
Trajectories that start at 
higher initial temperatures (lower $\beta_i$) 
remain substantially lower on Figure 3 throughout the entire 
evaporative trajectory, but
do eventually wind along the limit cycle at $\mu^*$. 
towards large $\beta$. Recall that in the active method we are actually
starting generally at smaller $\beta_i$ than shown.
\medskip
\centerline{ \psfig {figure=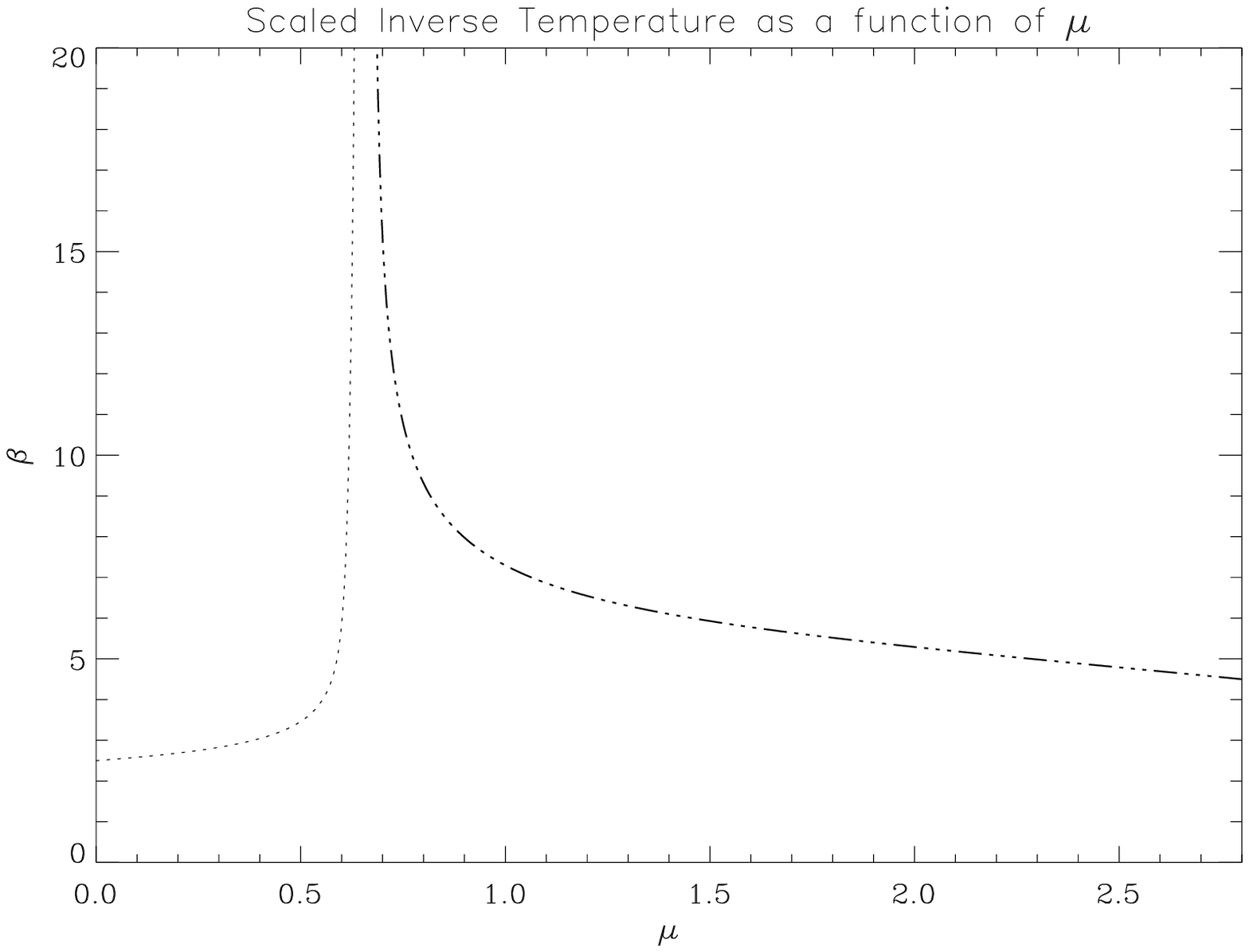,height=2.0in,angle=0}}
\smallskip
\centerline{\it Figure 3: Two Typical Cooling Trajectories, both starting at 
$\beta \sim 4$, from above and below $\mu^*$.}
\medskip
Curiously, note that for $\mu >> \mu^*$, 
the $\mu\beta$ product can actually initially {\it decrease}
along the evaporative trajectory. In the conclusion we comment on a heuristic 
way of understanding such counter-intuitive behavior. 
At any rate, it is clear from Figure 3 that achieving a condensed fermi system
requires evolving $\mu$ close to $\mu^*$. We now 
investigate some liberal bounds on  
how well evaporation can achieve that goal.

In Figure  3 we integrated Eq.(3) and Eq.(4) for $\beta(\mu)$ by eliminating
$N$, not taking into account that $N$ must always be decreasing!  We now 
study how far along these $\beta(\mu)$ trajectories 
we may progress evaporatively until 
we substantially run out of particles. 

Analytically, at low temperature, we find that the 
flow equation for $\mu$ near $\mu^*$ is independent of 
$\beta$ entirely, and reads, 
$$ {{{\rm d}\mu}\over{{\rm d}N}} = {{\pi^2}\over{6}} {{2-3\mu}\over{\mu^3}}
+\ldots
\eqno(7)$$
which can be integrated for all $\mu$ to read 
$$\biggl[ -{{8}\over{27}}\log(\mu-2/3) 
- \mu^2{{\mu+1}\over{3}} - {{4\mu}\over{9}} \biggr]_{\mu_i}^{\mu_f} = {{\pi^2}\over{2}} \Delta N
\eqno(8)$$
The number evaporated $\Delta N$ must of course
be less than the total number of particles in the trap. 
Using Eq.(1), we see that there are indeed stringent limits on 
the R.H.S. of Eq.(8). Since we know that 
appreciable cooling in this scheme doesn't occur until 
one is close to $\mu^*$(=2/3 at low temperature), we know that the 
best one can do is to evaporate all the particles in excess 
of the ground state at $\mu^*$. For the scenario in which we 
start at a $\mu$ above $\mu^*$ (for example, as may be created in the 
active method) this limits the RHS substantially. 
For $\mu<\mu^*$, the system evaporatively evolves towards 
$\mu^*$ {\it but can never reach it} because there are  simply not 
enough particles to evaporate. That is one reason to use the active method
in a phase of the cooling, since it can raise the initial 
$\mu$ above the $\mu^*$. However, raising $\mu$ by this means is
also self-limiting for two reasons.
First, if it is raised substantially above 1
too many particles are lost from the trap and there are too few remaining to 
evaporate back to degeneracy. 
Secondly, as described earlier, it reduces $\beta_i$ by the same factor 
it increases $\mu_i$, indicating the need to get even closer to $\mu^*$ 
to wind along the limit cycle and recover large $\beta$. 
We explain this in more detail 
quantitatively below. 

We graphically describe the consequences of 
Eq.(8). All the discussion here is in the ``best case'' scenario, in that
we imagine starting with a system at already relatively large $\beta$ 
(so Eq.(8) applies) 
and ask how well evaporative cooling can further increase
$\beta$ and thus $\beta\mu$. 
Figure 4 is a graph of Eq.(8). The dashed line represents a 
maximum possible RHS at that initial $\mu$. As per the preceding 
discussion we have plotted the contribution from the 
total number of particles 
for $\mu<\mu^*$, and plot the {\it excess only} for $\mu>\mu^*$. 
The light dotted trace is the LHS of Eq.(8). Thus, to estimate the 
maximum possible increase/decrease in $\mu$, start at some initial 
$\mu_i$, use the height of the dashed line to estimate how much of
a change in the height of the light dotted line you may achieve. 
The resulting position at that height on the light dotted curve 
then is an (over-)estimate of the 
largest $\mu$ achievable\footnote{$\dagger$}{Note that 
although ${{\partial N}\over{\partial \mu}}|_\beta$ is
always greater than zero, the total derivative
${{{\rm d} N}\over{{\rm d} \mu}}|<0$ 
for $\mu<\mu^*$ because ${{\partial N}\over{\partial \beta}}|_\mu
{{{\rm d} \beta}\over{{\rm d} \mu}}$
is negative.}
\medskip
\centerline{ \psfig {figure=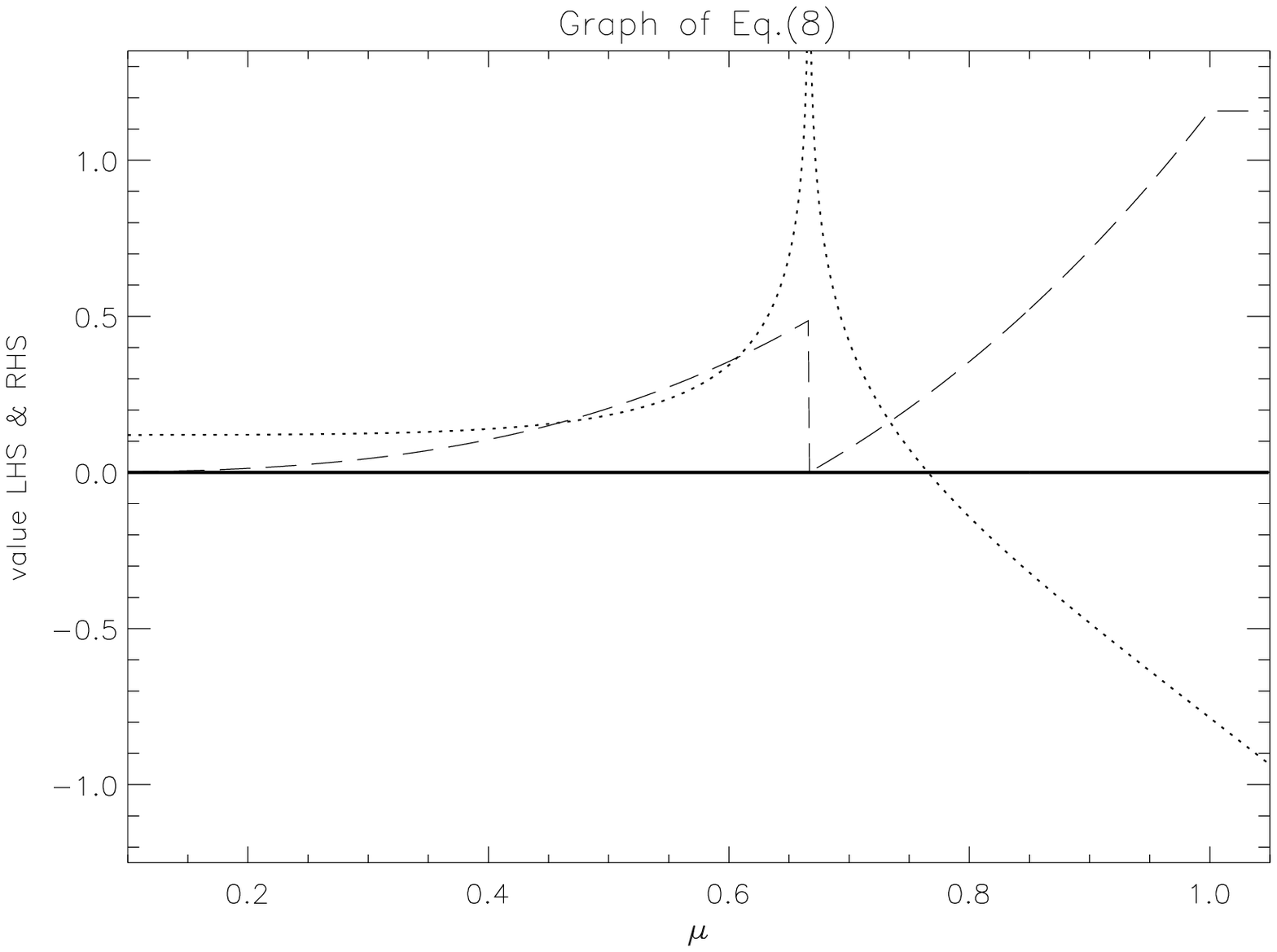,height=2.0in,angle=0}}
\smallskip
\centerline{\it Figure 4: Graph of Eq.(8)}
\medskip
Studying the graph indicates that the process allows one to 
get within perhaps $~10\%$ of $\mu^*$ at best. As a  very crude 
estimate, we can see that this occasions at most a roughly 3-fold increase
in $\beta$. Cooling from a typical active-method initial state of 
$(\beta,\mu) \sim (1,1)$, we find that one cannot achieve 
$\beta\mu$ products in excess of roughly 3. For such a gas, the 
typical occupation probability of the lowest energy 1-particle state
in the trap is roughly 90\%. Of course, measuring this occupation 
probability is not a good measure of quantum degeneracy 
because it is exponentially sensitive to $\beta\mu$. We 
prefer thinking about nearly degenerate 
quantum fermi systems as those with large $\beta\mu$ products
such as 
are found for nucleons in nuclei and 
electrons in typical metallic systems.
\bigskip
\noindent {\bf III: A Second Model:} We now study the scenario 
in which the evaporative thresholds for the 
two species are very different. 
We forgo a detailed quantitative analysis specific for this
case, and instead reduce to and reason from the 
simpler model we have already discussed. 
For simplicity we assume that the
masses and the trap potential of the two species are identical.
Thus, the quasi-equilibrium expectation values of $N_{a,b}$ and
$E_{a,b}$ are given by the obvious doubling- and indexing- of Eq.(1).  
However to include the fact that the evaporative 
threshold for species $a$ to be so much larger 
than that of species $b$ 
replace the upper limit '1' of the integrals in 
the Eq.(1) by '$\infty$'. We then scale $\beta$, $\mu_a$ and 
$\mu_b$ by 
the evaporative threshold of the $b$ species.
Subsequently $E_{evap}$ refers to the 
evaporative threshold for species $b$ only. 

The condition, ${\rm d}N_a=0$
now becomes a linear constraint in the space $(\mu_a,\mu_b,\beta)$.
All these variables are dimensionless, scaled by the 
appropriate factors of $E_{evap}$. 

That linear constraint reduces the evaporative evolution of this
two-component fermi system 
again to a two-dimensional dynamical system
({\it i.e.} at fixed $E_{evap}$ and $N_a$, $\mu_a$ is really a 
function of $\beta$). 
In particular, scattering events
that leads to evaporation of a particle of species $b$ at energy $E_{evap}$
will, on average, remove a net $E_a$ from the total energy of species
$a$ and $E_b$ from species $b$, where $E_a+E_b=E_{evap}$ and with 
ratio $E_a/E_b$ depending on 
$(\mu_a, \mu_b, \beta)$. In equations, ${\rm d}N_a=0$ implies that
$${{{\rm d} E_a}\over{{\rm d}\beta}} = -{{det(M_a)}\over{
{{\partial N_a}\over{\partial \mu_a}}|_{\beta}}}
\eqno(9)$$
(compare to Eq.(6))
where $det(M_a)$ is the determinant of the matrix of partial derivatives
for the $a$ system only. Note that this determinant $det(M_a)$ 
is now computed 
with the integrals extending to $\infty$, and so is singular at
high temperatures, but still looks like the rest of the graph in 
Figure 1 for low temperatures. The analogous function for the 
$b$ species, $det(M_b)$, is precisely the same as for Eq.(5) with Eq.(1)
(that is, using integration limits $[0:1]$). 

Energy conservation implies (in scaled dimensionless quantities) 
that the evaporative trajectory is along
$$  \biggl(\matrix{ {{\rm d}N_b} \cr 
{{\rm d}(E_a+E_b)} \cr 
}\biggr) = \biggl(\matrix{ {-1} \cr 
 {-1} \cr 
}\biggr) {\rm d}N_b
\eqno(10)$$
The differential relations between $E_b$, $N_b$, $\mu_b$ and
$\beta$ are exactly the same as for the matrix system 
analyzed in the first model. 
We now use Eq.(9) to rewrite the ${\rm d}E_b$ term in 
Eq.(10) in terms of ${\rm d}\beta$ and use it to re-write the 
two dimensional system for the evolution of $\mu_b$ and $\beta$ 
in terms of 
${\rm d}N_b$. Recall that $\mu_a$ also changes, but is 
given parametrically in terms of $\beta$ (and $N_a$, which is held fixed). 
We find, 
$$ {{\partial \beta}\over{\partial N_b}} = 
{{1}\over{det({\cal M})}} \biggl( -{{\partial E_b}\over{\partial \mu_b}}|_\beta
+{{\partial N_b}\over{\partial \mu_b}}|_{\beta}\biggr)
\eqno(11)$$
$$ {{\partial \mu_b}\over{\partial N_b}} = {{1}\over{det({\cal M})}}
\biggl( {{\partial E_b}\over{\partial \beta}}|_{\mu_b} - 
{{\partial N_b}\over{\partial \beta}}|_{\mu_b}
-{{det(M_a)}\over{ {{\partial N_a}\over{\partial \mu_a}}|_\beta}}
\biggr)
\eqno(12)$$
The denominator is
$$det({\cal M}) = det(M_b) + det(M_a) \biggl({{ {{\partial N_b}\over{\partial 
\mu_b}}|_\beta}\over{ {{\partial N_a}\over{\partial \mu_a}}|_\beta}}
\biggr)
\eqno(13)$$
and is again strictly positive. 

We can now use the intuition based on the first model discussed 
in detail above to constrain how well one can cool the $a$
species in this scenario. 
The measure of how close to the ground state we are for that 
species is again the product $\beta \mu_a$. 
Consider two possible initial high temperature 
limits: $N_a>>N_b$ and $N_a << N_b$. 
In the first case, $\mu_a$ is high but there are fewer
$b$ species to evaporate off, and by the analysis of the 
first model, we expect only a modest increase in $\beta$.
In the other extreme, $\mu_a$ 
starts off small, and so, although in principle you could 
carry off much heat, it's temperature 
will track the temperature of the $b$ species, which will 
again be limited by all the considerations discussed in the 
first model, and so the product $\beta \mu_a$ will start 
small and stay small. 
Thus, the most promising initial (high temperature) state is  
$N_a \sim N_b$ (that is, $\mu_a \sim \mu_b$).

Looking now at Eq.(12), we see that the 
limit cycle still exists and
occurs at a 
value $\mu_b^*$ {\it lower}
than in the first model we discussed. 
Also note that the $\beta$ evolution equation in 
this scenario differs only by the prefactor, $det({\cal M})$, 
from the first model we considered. 
Due to positivity of the individual $det(M_{a,b})$, 
we expect that factor to {\it reduce} the 
evolution of $\beta$ as compared with the first model. 
Indeed, in the 
high temperature limit the 
suppression is through powers of the ratio of the 
evaporative energy scales of the two species, but for 
low temperatures, $\det({\cal M}) \sim 2\det(M_b)$. 
Thus, as may have been expected due to the larger thermal 
inertia of the entire system compared with that of an individual species, 
the actually overall evaporative cooling efficiency 
is suppressed at low temperatures
relative to the first model we considered. 

This means that if we cool in the 
active mode, $\mu_b$ will be large initially and 
thus the temperature of the entire system will not drop 
dramatically as one evaporates. 
Efficient cooling could occur if we are able to get to 
$\mu \sim \mu^*$, but it is difficult to reach that regime
because, although we can evaporate {\it all} of species ``b''
(thereby having a continuous curve for the dashed curve in 
Figure 4 representing the upper bound for the RHS 
of Eq.(8)), in the low temperature limit the cooling of the 
whole system proceeds slower, and so the light dotted curve in 
Figure 4 is roughly twice as steep. Thus, the obstruction to 
reaching the quantum ground state in the two component system
in this regime (unequal $E_{evap}$)
can be understood from the considerations and 
qualitative behavior  of the first regime
(equal $E_{evap}$).

\bigskip
\noindent {\bf IV: Conclusion:} 
There exists a limit cycle in 
evaporative cooling a two component fermion system that has the consequence of 
severely limiting the approach to the quantum degenerate ground state. 
One heuristic way to understand this result is that since there is always a gap
between the putative fermi surface and the evaporative threshold, evaporation 
can actually ``heat up'' (that is, disorder) the distribution function 
at the fermi surface. This one way to understand the rather 
counter-intuitive finding that that starting at $\mu>>\mu^*$, the 
$\mu\beta$ product actually initially {\it decreases} 
during evaporation. 

There are many obvious reasons why the two component fermi system is so 
different than the bose system. From the point of view of the discussion 
here, the trapped bose gas starts with $\mu<<0$ and, 
analogous to what we have been discussing, evaporatively evolves towards
a limit cycle at some 
$\mu^*>0$. But, of course due to the singularity of the system at 
$\mu\rightarrow 0^-$, it never gets to $\mu^*$ but instead condenses. 

There have been many proposals for surmounting the difficulty of 
achieving a degenerate fermi ground state, and it is not the 
purpose of this note to review these many inventive ideas. They 
include condensing fermi-bose mixtures (a difficult technical feat)
$^{14,15,16,17,19}$ 
and using various perturbing fields on pure fermi systems$^{}$. 
What we suggest is that
many cooling techniques that evaporate fermions 
exclusively may be constrained in 
the same way as described above. 

We would like to end with a brief speculative
proposal for reaching lower $\beta\mu$ products in a trapped atomic 
fermi gas. Consider photoassociating fermi dimers
into states just below the trap single particle
ground state. With
fermions there is no stimulated
atomic channel back to the trap as would the case of 
photoassociating dimers from a bose condensate; instead  
pauli blocking and the fermi energy both push the 
system towards dimerization. The remaining fermions then scatter off the 
dimers. In a sense, photoassociating has 
enhanced the 
three-body collision rate, which, 
even for identical fermions, is not suppressed by statistics
at low energies. 
Every time a dimer breaks in collision, as long as 
the trapping potential is high enough,
the fermions go back into the trap; the net effect of creating and 
breaking dimers in this proposed scheme is to use the difference in the dimer 
pump beams to cool the fermion system 
``from below'' (near the single particle ground state of the trap) 
instead of evaporatively ``from above'' (that is, above the fermi surface). 
In that sense this scheme has the flavor of bose-fermi mixture
schemes, but might be simpler technically. 
Also, this cooling proposal does not a priori require a 
two fermion mixture, though we imagine photoassociating into a 
dimer composed  
of dissimilar fermions is likely to be easier than in identical ones.  
It remains to 
be seen whether such a technique can be practically implemented
in a polarized atomic fermi system.

Finally, it would be of great interest to compare the predictions 
of this simple evaporative model directly with experiment. One step 
in that direction is to generalize the model to include a time dependent  
$E_{evap}$ and trap lifetime effects.  
One use of such a direct comparison would be to further 
test how large quantum statistical effects are in 
current experiments which are far from degeneracy. 
Investigations of such ``dynamical'' effects
are underway but clearly beyond the ``kinematic'' scope and spirit 
of this note. 
\bigskip
\noindent{\bf V: Acknowledgments} It is a pleasure to thank D. S. Jin, 
R. Walsworth and W. Phillips for stimulating conversations and 
suggestions. We also thank J. Baltisberger for numerical work 
leading to Figure 1. 
This research was supported in part by Research Corporation 
Cottrell Science Award \#CC3943 and in part by the National Science 
Foundation under grants PHY 94-07194  and EPS-9874764.

\vfill
\eject
\ \ 

\centerline{\bf Bibliography}
\bigskip
\item{1.} M. H. Anderson, J. R. Ensher, M. R. Mathews, C. E. Weiman and 
E. A. Cornell, Science {\bf 269}, 198 (1995).
\medskip
\item{2.} K. B. Davis {\it et. al.}, Phys. Rev. Lett. {\bf 75}, 3969 (1995).
\medskip
\item{3.} C. C. Bradley, C. A. Sackett and R. G. Hulet, (to be published).
\medskip
\item{4.} I. Silvera, Physica 109 \& 110B (1982) 1499-1522. 
\medskip 
\item{5.} J. Oliva, Phys. Rev. B {\bf 38}, 8811 (1988) and Phys. Rev. 
B {\bf 39}, 4204 (1989).
\item{} D. A. Butts and D. Rokhsar, Phys. Rev A {\bf 55} 4346 (1997). 
\medskip
\item{6.} H. T. C. Stoof and M. Houbiers, cond-mat/9808171. 
\medskip
\item{7.} B. DeMarco and D. S. Jin,  Phys. Rev. A {\bf 58}, R4267 (1998).
\medskip
\item{8.} G. Brunn, Y. Castin, R. Dum, K. Burnett, cond-mat/9810013. 
\medskip
\item{9.} B. DeMarco, Bohm, Burke, Holland, D. S. Jin, {\it Phys. Rev. Lett.},
{\bf 82}:(21) 4208, (1999), cond-mat/9812350.
\medskip
\item{10.} C. J. Myatt, E. A. Burt, R. W. Ghrist, E. A. Cornell and 
C. E. Weiman, Phys. Rev. Lett. {\bf 78}, 586 (1997).
\medskip
\item{11.} G. Ferrari, {\it Phys. Rev. A}, {\bf 59}:(6) R4125, (1999),
cond-mat/9904162. 
\medskip
\item{12.} W. Geist, A. Indrizbecovic, M. Marinescu, T. A. B. Kennedy 
and L. Youm, cond-mat/9907222.
\medskip
\item{13.} D. S. Jin, Private Communication. 
\medskip
\item{14.} N. Nygaard and K. Moelmer, cond-mat/9901160.
\medskip
\item{15.} L. Vichi, M. Ingusclo, S. Stringari and G. M. Tino, 
{\it J. Phys. B - AT MOL OPT} {\bf 31}:(21) L899,(1998), cond-mat/9810115.
\medskip
\item{16.} M. Amoruso, A. Minguzzi, S. Stringari, M. P. Tosi and L. Vichi, 
{\it Eur. Phys. J. D.} {\bf 4}:(3) 261, (1998), cond-mat/9810210.
\medskip
\item{17.} T. Miyakawa, K. Oda, T. Suzuki and H. Yabu, cond-mat/9907009  
\medskip
\item{18.} L. Vichi and S. Stringari, cond-mat/9905154.
\medskip
\item{19.} W. Geist, L. You and T. A. B. Kennedy, {\it Phys. Rev. A} {\bf 59}:(2)
 1500, (1999).

\vfill
\par
\end